\documentclass[]{aastex61}
\usepackage{graphicx}
\usepackage{textgreek}
\usepackage{bm}

\def\insitu{in situ} 
\def\wind{\textit{Wind}}
\def\stereo{\textit{STEREO}}
\def\sta{\stereo\textit{ A}}
\def\stb{\stereo\textit{ B}}
\def\stab{\sta{ and \textit{B}}}

\newcommand\vex{\textit{VEX}}
\newcommand\soho{\textit{SOHO}}
\newcommand\ace{\textit{ACE}}
\def\sdo{\textit{SDO}}
\def\goes{\textit{GOES}}
\def\fixb{F$\beta$} 
\newcommand\kmps{\mbox{km s$^{-1}$}}

\newcommand\rsun{$R_\sun$}  

\submitted{\small Accepted by \textit{The Astrophysical Journal} on April 11, 2017; unproofed.}

\begin{document}
\title{Multi-spacecraft Observations
of the Coronal and Interplanetary Evolution
of a Solar Eruption {Associated with} Two Active Regions}

\author[0000-0001-8188-9013]{Huidong Hu}
\affil{State Key Laboratory of Space Weather,
    National Space Science Center,
    Chinese Academy of Sciences, Beijing 100190, China}
\affil{University of Chinese Academy of Sciences,
     Beijing 100049, China}
\author[0000-0002-3483-5909]{Ying D. Liu}
\affil{State Key Laboratory of Space Weather,
    National Space Science Center,
    Chinese Academy of Sciences, Beijing 100190, China}
\affil{University of Chinese Academy of Sciences,
     Beijing 100049, China}
\author[0000-0001-5205-1713]{Rui Wang}
\affil{State Key Laboratory of Space Weather,
    National Space Science Center,
    Chinese Academy of Sciences, Beijing 100190, China}
\author[0000-0002-4016-5710]{Xiaowei Zhao}
\affil{State Key Laboratory of Space Weather,
    National Space Science Center,
    Chinese Academy of Sciences, Beijing 100190, China}
\affil{University of Chinese Academy of Sciences,
     Beijing 100049, China}
\author[0000-0001-6306-3365]{Bei Zhu}
\affil{State Key Laboratory of Space Weather,
    National Space Science Center,
    Chinese Academy of Sciences, Beijing 100190, China}
\affil{University of Chinese Academy of Sciences,
     Beijing 100049, China}
\author[0000-0002-1509-1529]{Zhongwei Yang}
\affil{State Key Laboratory of Space Weather,
    National Space Science Center,
    Chinese Academy of Sciences, Beijing 100190, China}

 \correspondingauthor{Ying D. Liu}
 \email{liuxying@spaceweather.ac.cn}

\begin{abstract}
We investigate the coronal and interplanetary evolution
of a coronal mass ejection (CME) launched on 2010 September 4
from a source region {linking} two active regions (ARs) 11101 and 11103,
using extreme ultraviolet imaging, magnetogram, white-light
and \insitu{} observations from \sdo, \stereo, \soho, \vex{} and \wind.
A potential-field source-surface model is employed to examine
the configuration of the coronal magnetic field surrounding the source region.
The graduated cylindrical shell model
and a triangulation method are applied to determine the kinematics of the CME
in the corona and interplanetary space.
From the remote sensing and \insitu{} observations we obtain some key results:
(1) the CME was deflected {in both the eastward and southward directions}
in the low corona by the magnetic pressure from the two ARs
and possibly interacted with another ejection,
which caused that the CME arrived at \vex{}
that was longitudinally distant from the source region;
(2) although \vex{} was closer to the Sun,
the observed and derived CME arrival times
at \vex{} are {not earlier} than those at \wind,
which suggests the importance of determining both the frontal shape
and propagation direction of the CME in interplanetary space;
(3) the ICME was compressed in the radial direction
while the longitudinal transverse size was extended.
\end{abstract}


\section{Introduction} \label{sec:intro}
Coronal mass ejections (CMEs) are large expulsions of plasma
and rapid releases of magnetic field energy from the solar atmosphere.
They are called interplanetary CMEs (ICMEs) when propagating in interplanetary space,
which are one of the major disturbances in the solar wind.
ICMEs are also significant factors inducing geomagnetic activities
when they impact the Earth.
From the low corona to interplanetary space,
a CME may experience non-radial motion,
interact with other CMEs and be deformed
due to the interaction with the ambient solar wind,
involving plenty of physical processes.

The configuration of the magnetic field surrounding the source region
as well as the interaction with other structures in the solar wind
are important for estimating the propagation direction of a CME
and thus crucial for space weather forecasting.
Previous coronagraph observations suggested
that latitudinal deflections of CMEs were controlled
by the background coronal magnetic field and the overall flow near solar minimum
\citep{mhc1986JGR},
which was confirmed by the result
that CMEs near solar minimum
were systematically deflected toward lower latitudes
while CMEs during solar maximum could deviate
toward either the pole or equator
\citep[e.g.,][]{cb2004AA,kpv2009AnGeo}.
Some CMEs were found to be deflected toward the weak magnetic field region
associated with the streamer belt or the heliospheric current sheet
\citep[e.g.,][]{xsg2009SoPh,koe2013ApJ,lpo2015}, while
\citet{xsg2009SoPh} reported that
fast CMEs could be deflected away from the streamer belt.
The coronal hole with strong open magnetic field
is one important factor that can divert a CME
\citep[e.g.,][]{gyk2004JGRA,gym2005GeoRL,gmx2009JGRA,xsg2009SoPh,pmv2013SoPh}.
\citet{mrf2015natco} and \citet{wld2015apj}
found that the strong magnetic field of a sunspot
in an active region (AR)
and the configurations of the magnetic field surrounding the source region
play a significant role in channeling the CME.
Deflection of CMEs in the corona is likely to occur
within a few solar radii from the Sun
\citep[e.g.,][]{ivk2014SoPh,koe2015ApJ,wld2015apj},
while CME-CME interaction
may change the propagation direction of a CME in interplanetary space
\citep[e.g.,][]{gyk2001ApJ,llm2012,llk2014NatCo,lfd2012ApJ}.

A plethora of efforts have been made to understand the morphology of the CME
in the corona and interplanetary space,
which is crucial to determine the arrival time of the CME at a planet.
\citet{thv2006ApJ,tvh2009SoPh} presented the graduated cylindrical shell (GCS) model
employing a croissant-like empirical flux-rope structure
to describe the morphology of the CME near the Sun.
A compact structure was used to derive CME kinematics
\citep[e.g.,][]{sww1999}.
A circle attached to the Sun,
called the harmonic mean method,
was also applied to determine the kinematics in the inner heliosphere
\citep[e.g.,][]{lvr2009AnGeo}.
A self-similar expansion (SSE) geometry
with an expanding circle in the ecliptic plane
was adopted to describe the CME front
\citep{lhr2010,dhp2012ApJ,md2013SoPh}.
\citet{mrf2015natco} and \citet{rmi2016ApJ} introduced an elliptical shape
with a constant angular width to forecast CME arrival time and speed
based on single-viewpoint imaging observations.
The readers are directed to
\citet{ltl2010apj} for discussions of some of the geometries.
However, a CME could be deformed by interacting with the ambient solar wind
\citep[e.g.,][]{rlm2003,orz2004,mgr2004,lrb2006JGRA,sor2010ApJ},
as well as structures in the solar wind
\citep[e.g.,][]{lvr2009AnGeo,llm2012,lhw2015,ksk2015GeoRL}.
These interactions can alter their plasma and magnetic field characteristics
as well as morphologies in interplanetary space.

In this manuscript we investigate the coronal and interplanetary evolution
of a CME launched from a source region {linking two ARs.}
The eruption may occur as a result of {magnetic coupling}
of the two ARs, but it is different from the cases of sympathetic eruptions
\citep[e.g.,][]{st2011JGRA,jsc2016ApJ}.
The remote sensing and \insitu{} observations
from \sdo, \stereo, \soho, \vex{} and \wind{} are used in this study.
The positions of the spacecraft and planets during the event
are indicated in Figure \ref{fig:pos}.
\sta{} was ${\sim}$81{\degr} ahead of the Earth
{and was $\sim$20{\degr} west of the source region ($\sim$W62{\degr}),}
and \stb{} was ${\sim}$74{\degr} behind the Earth.
\vex{} (Venus) was $\sim$0.73 au from the Sun and $\sim$31{\degr} east of the Earth.
The spacecraft are at positions {that are advantageous for investigating}
the kinematics and the structure of the CME.
The study illustrates
{how the magnetic field of the two ARs
on one side of the source region}
change the motion of the CME near the Sun.
The multi-spacecraft \insitu{} measurements and remote sensing observations
{place constraints} on the structure and the kinematics of the CME,
{and demonstrate how the frontal shape and propagation direction
determine the arrival time of the CME.
Furthermore,} multiple spacecraft at longitudinally separated positions
reveal the longitudinal transverse size of the ICME,
which is important for understanding the structure and distortion
of the ICME in the solar wind.
The evolution of the CME in the corona and interplanetary space is presented
in Sections \ref{sec:corona} and \ref{sec:interplanet}, respectively.
We conclude and discuss the results in Section \ref{sec:conclusion}.
The results of this work shed light on
the kinematics and morphological evolution of the CME
in the inner heliosphere as well as space weather forecasting.

\section{Evolution in the Corona} \label{sec:corona}
The CME was launched around 14:30 UT on 2010 September 4
in the rising phase of solar cycle 24,
associated with two roughly straight flare ribbons
{on the eastern edges of
two ARs 11101 (W67{\degr}N11{\degr}) and 11103 (W69{\degr}N26{\degr}),
as shown in the right panel of Figure \ref{fig:euv}.
Figure \ref{fig:euv} displays the extreme ultraviolet (EUV) images
and magnetograms about the source region from \sdo{} and \sta.
In the magnetograms,
the white (dark) regions represent the positive (negative) magnetic field regions,
and the locations of ARs 11103, 11101 and 11102
are marked with ``A'', ``B'' and ``C'', respectively.}
According to the EUV observations of \sdo{/AIA
\citep{lta_aia2012} and \sta/EUVI
\citep{hmv_secchi2008},
the ribbons became visible at wavelength 304 {\AA} at about 14:16 UT
and separated eastward and westward, respectively.
From the \sdo/HMI
\citep{ssb_hmi2012} {magnetogram}
overlaid with the intensity contour
of the AIA 304 {\AA} image,
the magnetic field corresponding to the eastern ribbon is positive
and the magnetic field for the western ribbon is negative.
The neutral line between the two ribbons
is along the meridian and at a longitude of {$\sim$W62{\degr}}
($\sim$77{\degr} in Carrington longitudes).
{Viewing from the viewpoint of \sta,}
a potential-field source-surface (PFSS) model
{reveals an overlying loop structure above the two flare ribbons
as shown in the lower right panel of Figure \ref{fig:euv}.}
The associated loop brightening during the eruption
was also observed at wavelength 195 {\AA} by {\sta}/EUVI
(not shown here).
A hot channel, which is considered as an erupting flux rope
\citep[e.g.,][]{zcd2012NatCo,cdz2014ApJ},
is not observed in the EUV images.
{However, an eruption with a two-ribbon flare and loop brightening
is considered evidence of an erupting flux rope
\citep[e.g.,][]{alm2000ApJ,qwc2004ApJ}.
In our case, an erupting flux rope could be inferred
from the flare ribbons and the loop brightening,
which was lying over the neutral line between the two ribbons.}
The erupting flux rope probably was of low twist
because the magnetic field near the neutral line
was relatively weak and not strongly sheared.
As can be {inferred} from the combined EUV and HMI observations
in the upper right panel of Figure \ref{fig:euv},
{the erupting flux rope likely extended}
from the positive polarity region of AR 11103
{(marked with ``A'')} in the north
to the negative polarity region of AR 11101
{(marked with ``B'')} in the south.
{The source region of the CME was expected to be between the two ribbons
and adjacent to the eastern edges of the two ARs.}
{ARs 11103 and 11101 are separated by $\sim$15{\degr} in latitudes
but may be connected through the erupting flux rope
that is under the PFFS model revealed loop structure.}
The maximum magnitude of the HMI measured magnetic field
in AR 11101 is about 2100 gauss,
and the one in AR 11103 is about 1600 gauss.
\citet{mrf2015natco} and \citet{wld2015apj} reported
that the strong magnetic field {(over 3000 gauss)} of a nearby sunspot
could change the propagation direction of the CME near the Sun.
In our case,
{although the magnetic field is not so high,
the asymmetric magnetic field configuration close to the source region}
may also interfere the motion of the CME in the corona.

{Figure \ref{fig:pfss_be} shows the magnetic energy density distribution
surrounding the CME source region at heights of 1.75 and 2.25 \rsun{}.
The magnetic energy density $w_ \mathrm B$ is calculated from
$ w_\mathrm B = {\mathbf{B}^2}/{2\mu_0}$,
where the magnetic field $\mathbf B $ is from the PFSS extrapolation
and $\mu_0$ is the vacuum permeability.
In Figure \ref{fig:pfss_be},
the two vertical dashed lines indicate the approximate positions
of the two ribbons, and the source region is between them.
For the sake of simplicity, we use contours of the magnetic energy density
to infer the general direction of the gradient descent.
As displayed in the upper panel of Figure \ref{fig:pfss_be},
the contour reveals that the gradient descent of the magnetic energy density
near the source region is generally eastward
at the height of 1.75 \rsun{}.
As shown in the lower panel of the figure,
the gradient descent near the source region
at the height of 2.25 \rsun{} has a southward component.
ARs 11103, 11101 and 11102 marked with diamond symbols in Figure \ref{fig:pfss_be}
are generally in the regions of higher magnetic energy density
on the west side of the source region.
This implies that
ARs 11103, 11101 and possibly 11102
caused an asymmetric magnetic field configuration
near the source region.
Previous studies suggested that
an asymmetric configuration of the coronal magnetic field
could deflect a CME toward the weak magnetic field region
\citep[e.g.,][]{mhc1986JGR,kpv2009AnGeo,koe2015ApJ,wld2015apj}.
We use the GCS model developed by
\citet{thv2006ApJ,tvh2009SoPh}
to investigate the propagation direction near the Sun,
and find that the propagation direction of the CME at 6.15 \rsun{}
(marked with a square in Figure \ref{fig:pfss_be})
is in the southeast of the source region,
which is consistent with the general direction
of the gradient descent of the magnetic energy density.
The GCS modeling will be detailed in the text below.}

{The GCS model can determine} the height, direction and other parameters
of {the CME} in the corona based on the coronagraph observations
from \stereo/SECCHI
\citep{hmv_secchi2008} and \soho/LASCO
\citep{dfp1995soho}.
Two coronagraphs are on board each \stereo{} spacecraft
\citep{kkd_stereo2008},
of which COR1 has a field of view (FOV) of 1.5--4 \rsun{}
and COR2 has a FOV of 2.5--15 \rsun{}
\citep{hmv_secchi2008}.
Coronagraph C2 on board \soho{} has a FOV of 1.5--6 \rsun{}
and C3 has a FOV of 3--30 \rsun{}
\citep{dfp1995soho}.
{Multi-spacecraft coronagraph observations from three viewpoints
place strong constraints on the parameters of the GCS model.}
The CME appeared in the FOV of \stb{/COR1} at 15:05 UT,
\soho/C2 (from position angle PA $\sim$270{\degr}) at 15:12 UT,
and \sta{/COR1} at 15:37 UT.
Running difference coronagraph images from \soho, {\stab}
as shown in Figure \ref{fig:gcs}
are used to fit the GCS model parameters of the CME.
{In every two rows of Figure \ref{fig:gcs},
two sets of identical running difference images are presented,
where the GCS modeled CME (green grids) are superimposed on the second set of images.}
In the running difference images of the top two rows of {the figure,}
we see that the CME was about to enter the FOV of \sta/COR1
at 15:25 UT when it was clearly observed by \stb/COR1 and \soho/C2.
This strongly restricts {the CME propagation direction}
at that moment with the GCS model,
because an {underestimated angle
between the Sun-Earth line and the propagation direction}
will bring the modeled CME (the green grids) into the FOV of \sta/COR1 then.
In the middle two rows of Figure \ref{fig:gcs},
the running difference image of \soho/C2 shows
{a loop-like structure in the northwest of the CME.
This structure may be a substructure of the CME,
and it is not fitted during the GCS model.
However, the possibility that the loop-like structure
is of another CME cannot be excluded.} The front of the CME was distorted
and the southern part moved faster than the part above the ecliptic plane
as displayed in the images of the bottom two rows of Figure \ref{fig:gcs}.
The distortion cannot be fitted by changing the tilt-angle parameter of the GCS model.
Only the part of the CME front near and above the ecliptic plane is fitted
by the GCS model when the distortion is significant.
{The distortion may be caused by a streamer
interacting with the southern part of the CME from the behind,
which can be seen in \stereo/COR2 white-light observations (not shown here).
At each time,
tilt-angle, aspect-ratio, half-angle and latitude parameters of the GCS model
are fixed while only the longitude and height parameters are fitted.}

{The fitted parameters of the GCS model
and their corresponding times are given in Table \ref{tab:gcs}.
The propagation direction of the CME}
at height 3.00 \rsun{} obtained from the GCS model
{was 60\fdg4 west of the Sun-Earth line.
It was close to the longitude ($\sim$W62\degr{})}
of the source region estimated from the EUV and HMI observations.
{The angle between the propagation direction and the Sun-Earth line
decreased remarkably to 48\fdg1}
when the CME height reached 6.15 \rsun{},
and then gradually changed to {45\fdg8} when the height was 17.57 \rsun.
During the whole GCS process,
the tilt-angle parameter is set to 90{\degr}
that is consistent with the observed meridian-aligned flare ribbons.
The constant tilt angle shows that the CME did not apparently rotate near the Sun.
The other three parameters are slightly adjusted according to the data times,
and the changes of aspect ratio and half angle
could be explained by the expansion of the flux rope.
{\citet{zbj2012ApJ} reported that
a CME could be deflected latitudinally about 15\degr{}
toward the equator within $\sim$3.5 \rsun{}.
The roughly constant latitude (around 0\degr{}) in Table \ref{tab:gcs}
indicates that the CME in our manuscript had been diverted over 10\degr{} southward
from the location of source region
($\sim$W62\degr{}N15\degr{} as indicated with the two vertical lines in Figure \ref{fig:pfss_be})
before reaching the height of 3 \rsun{}.
The GCS modeling results also show that
the CME was deflected over 10\degr{} eastward at the height of 6.15 \rsun.
The CME propagation direction at 6.15 \rsun{} is roughly
in the southeast of the source region,
which is near the regions of lower magnetic energy density
as shown in Figure \ref{fig:pfss_be}.
The deflection direction of the CME in the low corona
is roughly consistent with the distribution of the magnetic energy density.
The direction variation of the CME revealed by the GCS model
is minor above 6.15 \rsun{} as shown in Table \ref{tab:gcs},
which is consistent with that
the rate of CME deflection caused by the coronal magnetic field
is more significant within a few solar radii
\citep[e.g.,][]{ivk2014SoPh,koe2015ApJ}.}

\section{Properties in Interplanetary Space} \label{sec:interplanet}
The interplanetary kinematics of {the CME} is derived through
a geometric triangulation technique initially proposed by
\citet{ldl2010apjl}
based on stereoscopic wide-angle imaging observations from {\stereo}.
The triangulation method first utilizes
a {Fixed $\beta$ ({\fixb})} geometry for the CME front,
which assumes a relatively compact structure
\citep{sww1999,kw2007}
simultaneously observed by two separated spacecraft.
Later, \citet{lhr2010} and \citet{ltl2010apj}
incorporated a
{harmonic mean (HM)} approximation,
which assumes that the CME front
is a sphere attached to the Sun
and tangent to the lines of sight of the two spacecraft.
The triangulation concept has proved to work well
in acquiring CME interplanetary kinematics
and in connecting remote sensing observations with \insitu{} signatures
\citep[e.g.,][]{ldl2010apjl,ltl2010apj,lll2013,lhl2016,lhr2010,mtr2010,
hdm2012,dpt2013ApJ,hlw2016ApJ}.
The triangulation technique with \fixb{} and HM approximations
is detailed and discussed in
\citet{ltl2010apj,lll2013,lhl2016}.

The elongations and times as inputs of the triangulation method
are from the observations of \stereo{} COR2, HI1 and HI2.
The angular FOV is 0\fdg7--4{\degr} around the Sun for COR2,
$20{\degr} \times 20{\degr}$ centered at 14{\degr} from the center of the Sun for HI1,
and 70{\degr} in diameter centered at 53\fdg7 for HI2
\citep{hmv_secchi2008}.
The left panel of Figure \ref{fig:hi_jmap} depicts
the running difference images of the CME in interplanetary space
taken by HI1 and HI2 of \stab.
{The feature near the northern edge of the CME
in the HI1 images might be the substructure of the CME
that is also observed by \soho/C2.
We cannot rule out the possibility that the feature is of another CME.
The northern front of the CME was distorted
and the southern part of the front was advancing the other part.}
Through stacking the running difference intensities
of COR2, HI1, and HI2 within a slit along the ecliptic plane,
we construct two time-elongation maps
\citep[J-maps, e.g.,][]{sww1999}
for the CME as shown in the right panel of Figure \ref{fig:hi_jmap}.
{The red curves in the maps represent
the elongation angles of the CME front near the ecliptic plane
observed from the viewpoints of \stab{}, respectively.}
\sta{} tracked the CME out to $\sim$50{\degr}
while \stb{} followed the CME out to $\sim$40\degr,
{as indicated by the red curves.}
The elongations and times are extracted
and used as inputs of the triangulation method.
{The CME kinematics along the ecliptic plane
derived from the \fixb{} and HM triangulations
will be compared with the \insitu{} observations.}
The trajectories of the CME
derived from the triangulation with the {\fixb} and HM approximations,
as well as the positions of the spacecraft
in the ecliptic plane during the event
are displayed in Figure \ref{fig:pos}.
{The black diamonds and red crosses in Figure \ref{fig:pos}
represent the trajectory of the CME apex in the ecliptic plane
derived from the \fixb{} and HM triangulations, respectively.
The green and blue circles represent the sizes
of the HM assumed spherical front
when the CME arrives at \wind{} and \vex{}, respectively.
The arrows indicate the directions of the apex
when the CME arrives the two spacecraft.}

{The propagation direction, height and speed of the CME}
in the ecliptic plane derived from the triangulation method
and the kinematics in the corona obtained from the GCS model
are presented in Figure \ref{fig:prop}.
{In the top panel of the figure,
the derived propagation direction is given in degrees from the Sun-Earth line.}
The direction angle acquired from the HM triangulation
is roughly twice the one from the {\fixb} triangulation and more variational,
which is noticed in previous studies
\citep{lhr2010,lll2013,lhl2016}.
Near the Sun the longitude is overestimated by the HM approximation
which is probably a result of the overestimated CME size as discussed in
\citet{lll2013,lhl2016}.
Both the GCS model and the \fixb{} triangulation reveal that
the CME was propagating {to} the east of the source region,
and the direction changed dramatically toward the east in the low corona.
The direction angles from the GCS model are
between those from the \fixb{} and HM approximations.
The direction angles from the GCS model
are closer to those from the \fixb{} approximation at lower heights,
and have a trend to approach to those from the HM approximation at larger heights.
The CME was deflected from about 60\degr{} to about 50\degr{}
west of the Sun-Earth line below $\sim$6 \rsun.
{The difference between the propagation directions
before and after the data gap in the top panel of Figure \ref{fig:prop}
implies that the CME} may continue to turn eastward in interplanetary space.
The final propagation direction angle before reaching the Earth
is likely around $20\degr$ west of the Sun-Earth line,
to which both the \fixb{} and HM triangulations seem to converge.
Given these variations in the propagation direction,
{the CME possibly arrived at Venus (\vex)
that was longitudinally $\sim$90\degr{} distant from the source region
(see the text below).
There were other ejections launched near AR 11103 and 11102 on September 4,
which were all on the west side of the CME of interest.
One of these ejections may interact with the CME,}
which is likely to be the reason for
the direction change of the CME in interplanetary space.
{As shown in the middle and bottom panels of Figure \ref{fig:prop},}
the GCS model and the triangulation method
give consistent distances and speeds of the CME
below $\sim$20 \rsun.
The distances and speeds from the \fixb{} approximation
slightly deviate from those from the HM approximation beyond $\sim$130 \rsun,
which is probably due to the limitation of the geometric assumption
for the \fixb{} approximation at larger distances as discussed in
\citet{lll2013,lhl2016}.
Below $\sim$30 \rsun{},
the CME was gradually accelerated to about 600 \kmps.
Due to the data gap
we cannot tell if there is a deceleration process,
but the speeds beyond $\sim$60 \rsun{} do not show much variation
($\sim$620 \kmps{} for \fixb{} approximation
and $\sim$530 \kmps{} for HM approximation
as marked in Figure \ref{fig:pos}).

{Figure \ref{fig:vex} shows the solar wind magnetic field measurements
from the \vex{} magnetometer
\citep{zbd2006vexmag},
where the ICME
(with an interval of $\sim$7.5 hr from 11:04 UT to 18:33 UT on September 8)
is reported by
\citet{gf2016solphys}.
The high fluctuations of the magnetic field at lower altitudes
was probably caused by the induced magnetosphere of Venus.
\vex{} was} at a distance of $\sim$0.73 au
with a longitude of $\sim$$-31$\degr{}
that was over 90{\degr} away from the source region.
{The distance data range of the CME
derived from the observed elongations using the triangulation method
is not large enough to estimate the arrival times of the CME front
at \vex{} and \wind{}.}
Assuming a fixed propagation direction
($\sim$17\degr{} {from the Sun-Earth line} for \fixb{} and $\sim$24\degr{} for HM)
outside the data range,
we use a linear fit of the distances {determined by the triangulations}
to estimate the arrival times of the CME at \vex{} and {\wind}.
The HM triangulation suggests that
{the eastern flank of the CME
(the eastern part of the blue circle attached to the Sun in Figure \ref{fig:pos})}
arrived at \vex{} around 15:53 UT on September 8,
about 5 hr later than {the start time of the ICME observed at \vex{},}
with a radial speed at \vex{} of $\sim$310 \kmps.
Assuming that the compact structure for the \fixb{} approximation is the dominant apex
and the CME front is also a circle attached to the Sun like the HM approximation,
we estimate the arrival time at \vex{},
using the \fixb{} derived directions and distances,
to be 20:52 UT on September 7,
about 14 hr earlier than {the observed ICME start time.
The HM triangulation has well predicted the arrival times of ICMEs
in comparison to the \insitu{} observed start times of the ICMEs
\citep[e.g.,][]{ltl2010apj,lhl2016}.
In our case,
the small difference between the HM triangulation derived arrival times
and the ICME start time
suggests that the \vex{} observed ICME shown in Figure \ref{fig:vex}
is the interplanetary counterpart of the CME of interest.}

The associated \insitu{} measurements at \wind{} {during the event}
are displayed in Figure \ref{fig:wind}.
{Two ICME-like structures are indicated with ``I1'' and ``I2''
in the figure, one of which may be associated with the CME of interest.}
Although the alpha-to-proton density ratio is below the threshold for a typical ICME
\citep[e.g.,][]{rc2004JGRA,lrb2005PSS},
{the duration of structure I1}
($\sim$6 hr from 19:42 UT on September 7 to 1:39 UT on September 8)
can be determined by a combination of the increase of the alpha-to-proton density ratio,
the low proton temperature
\citep{rc1995JGR} and the magnetic field profile.
{Structure I2 has an interval of $\sim$15.5 hr
(from 10:15 UT on September 8 to 1:52 UT on September 9),
whose trailing edge may be ambiguous.}
The HM approximation determines the arrival time of the CME at \wind{}
to around 03:23 UT on September 8 and the estimated radial speed is $\sim$490 \kmps.
The \fixb{} approximation gives an arrival time around 18:17 UT on September 7.
{The HM estimated arrival time is $\sim$8 hr later
than the start time of structure I1,
and is $\sim$7 hr earlier than the start time of I2.
The \fixb{} estimated arrival time is $\sim$1 hr earlier
than the start time of I1,
and is $\sim$16 hr earlier than the start time of I2.
Given the small difference between the estimated and observed arrival times,
it is difficult to determine which of the two ICME-like structures (I1 and I2)
is associated with the CME of interest.
There was a brief structure (from 12:18 UT to 16:26 UT on September 9) after structure I2,
which is likely irrelevant to the CME of interest because of its late start time.}

{Figure \ref{fig:sta} shows the \insitu{} measurements
at \sta{} during the event.}
On September 7 \sta{} observed a shock leading an ICME
with the arrival time of 8:29 UT.
{The ICME interval is from around 13:00 UT on September 7
to around 9:45 UT on September 8,
which is determined from the combination of the low-temperature region
and magnetic field profile.
Both the shock arrival time and the ICME start time are
over 80 (60) hr earlier than the \fixb{} (HM) estimated arrival time.
Although the triangulation assumed shape
may deviate from the real geometry of the CME,
the large difference between the estimated and observed arrival times
can hardly associate the \sta{} observed ICME with the CME of interest.
This is also consistent with the statement in the ICME catalog of
\citet{gf2016solphys}
that no \insitu{} ICME signatures observed at \sta{}
were likely to be associated with the ICME at \vex{} in Figure \ref{fig:vex}.
A couple of ejections were launched from source regions
longitudinally close to \sta{} on September 4.
The \sta{} observed ICME might be associated with one of these ejections,
which might interact with the CME of interest
and caused the latter to continue turning eastward in interplanetary space.}

The {HM approximation} estimated arrival times
of the CME at \vex{} and \wind{}
are generally consistent with the observed,
{even though the ICME at \wind{} cannot be uniquely
associated with the CME of interest.}
This seems to suggest that the assumed spherical geometry
is applicable to {the eastern part of the CME front} in this case.
The \insitu{} measurements and the triangulation method both reveal that
{the arrival time of the CME at Venus (\vex)
is not earlier than that at the Earth (\wind),}
although the Sun-Venus distance ($\sim$0.73 au) was less than the Sun-Earth distance.
This is exactly what is shown in Figure \ref{fig:pos}:
the expanding eastern flank of the {assumed} CME circle touches \vex{}
after the front reaches \wind{}.
Therefore the arrival times of the CME
at the two spacecraft with different heliocentric distances
can be determined by the joint effect of the propagation direction
and the assumed expanding circular shape.
{However, it is unclear whether the assumed shape
could describe the geometry of the western part of the CME
because of the lack of consistent \insitu{} measurement of the western part.}
Assuming that the CME front is symmetric
with respect to the final propagation direction,
we estimate the longitudinal angular width of the CME front
to be {probably no less than 100{\degr}
given that the angular distance between \vex{}
and the propagation direction is $\sim$50\degr{}.}
Let the semi-perimeter of the circle
{(the leading edge of the flux rope)}
in the HM approximation
represent the longitudinal transverse size,
and we estimate the size to be about 1.7 au
when the CME arrived at \wind.
{The radial widths of structures I1 and I2 at \wind{}
(as indicated in Figure \ref{fig:wind}),}
obtained from the HM estimated radial speeds multiplied by the durations,
{are $\sim$0.07 au and $\sim$0.18 au, respectively.
Both the radial widths of structures I1 and I2 are far less than
the transverse size $\sim$1.7 au.}
Note that the flux rope is generally perpendicular to the ecliptic plane
as can be seen from Section \ref{sec:corona},
{so the transverse size is approximately equal to the transverse size
of the cross section of the flux rope in the ecliptic plane.}
This result suggests that the ICME structure
was extremely compressed in the radial direction
and extended in the longitudinal direction in interplanetary space.

\section{Conclusions and Discussions} \label{sec:conclusion}
We have investigated a CME launched on 2010 September 4
from a source region {connecting} AR 11101 and AR 11103,
based on EUV, magnetogram, white-light and \insitu{} observations
from \sdo, \stereo, \soho, \vex{} and \wind.
The PFSS model is used to examine the magnetic field environment
near the source region.
The GCS model and the triangulation method are applied
to analyze the kinematics and structure in the corona and interplanetary space.
Below we summarize the results and discuss their crucial implications for understanding:
1) how the propagation direction of the CME in the low corona
{is changed by the asymmetric magnetic field configuration
caused by the ARs on one side of the source region;}
2) the determination of both the CME frontal shape
and the propagation direction in space weather forecasting; and
3) {the deformation of the ICME structure.}
\begin{enumerate}
  \item The CME was deflected in the low corona
  by the magnetic pressure from the two nearby ARs in the west,
  and possibly interacted with another CME in interplanetary space.
  The erupting flux rope originated
  from the positive and negative polarity regions
  belonging to the two different ARs.
  With all the sunspots on the west side,
  the source region was adjacent to the eastern edges of the two ARs,
  which caused an asymmetric magnetic field configuration on the two sides of the eruption
  and resulted in a deflection of the CME in the low corona.
  {The deflection direction revealed by the GCS model
  is consistent with the general direction of the gradient descent
  of the magnetic energy density derived from the PFSS model.}
  This suggests that a CME originated from the edge of an AR is likely to be deflected
  toward a non-radial motion
  due to the asymmetric background magnetic field configuration.
  This is similar to the case of
  \citet{mrf2015natco} and \citet{wld2015apj}
  where the strong magnetic field on one side of the CME source region
  changed the propagation direction of the CME.
  However, a CME-CME interaction is possibly involved in the change of propagation direction.
  {Given that there were other ejections with close launch times
  and source regions to those of the CME of interest,
  the CME possibly interacted with one of these ejections during the propagation.
  The deflection and interaction may be the reason that caused the CME
  to reach \vex{} that was longitudinally distant from the source region.}
  CME-CME interaction is not unusual
  and may change the propagation direction and 1-au property of a CME
  \citep[e.g.,][]{gyk2001ApJ,lvr2009AnGeo,lfd2012ApJ,llm2012,llk2014NatCo,lhw2015}.
  This illustrates the necessity of considering
  both the background magnetic field condition
  and successive eruptions near the source region
  in determining the propagation direction of a CME.

  \item It is important to determine both the frontal shape and the propagation direction
  of the CME in interplanetary space for space weather forecasting.
  {Although the associated ICME at \wind{} cannot be uniquely identified
  from the two plausible structures,}
  the HM triangulation with a circular geometry in the ecliptic plane
  can give arrival times that are less than 8 hr
  {different from} observed at \vex{} and \wind{}.
  Both the observed and the triangulation estimated arrival times of the CME
  {at \vex{} ($\sim$0.73 au) are not earlier
  than those at \wind{} ($\sim$1 au),}
  although \vex{} was closer to the Sun than \wind.
  These results suggest that the simple expanding sphere attached to the Sun
  may describe the shape of {the eastern} CME front in the inner heliosphere.
  {However, the spherical shape is only consistent with
  \insitu{} measurements at two points,
  and there is no consistent observations of the western part of the CME.
  The real geometry of the CME front is not strongly constrained.
  Furthermore,} the CME could be distorted {to a complex geometry} so that
  {it arrived at \vex{} later.}
  Also note that the HM triangulation gives a misleading propagation direction near the Sun
  and large variations in the propagation direction.
  The CME touched \vex{} with its eastern flank in this case,
  and the flank of a CME could also trigger a geomagnetic storm as reported by
  \citet{mtr2010}.
  This indicates the importance of determining both the CME frontal shape
  and propagation direction in forecasting geomagnetic activity.

  \item The structure of the CME in interplanetary space
  was compressed radially and extended longitudinally,
  {leading to a large transverse size of the flux-rope cross section
  in the ecliptic plane.}
  Estimate of the transverse size and deformation of the ICME structure
  in the inner heliosphere could be helpful
  to improve our knowledge of the interaction between the ICME and the solar wind
  as well as space weather forecasting.
  For example, the transverse size can increase the contact area of the interaction
  between an ICME and the solar wind
  and the probability of arriving at the Earth,
  while the deformation could alter
  the interplanetary plasma and magnetic field properties of an ICME near the Earth.
  Observations from the \textit{Solar Orbiter}
  \citep{mms2013SoPhSolarOrbiter}
  out of the ecliptic plane can be of great advantage
  to study the kinematics and geometric evolution of CMEs,
  as well as their interactions
  with other structures in the solar wind.
\end{enumerate}

\acknowledgments
The research was supported by the Recruitment Program
of Global Experts of China, NSFC under grant 41374173
and the Specialized Research Fund for State Key Laboratories of China.
We acknowledge the use of data
from {\stereo, \sdo, \soho, \vex, \wind, \ace{} and \goes}.

\clearpage
\begin{deluxetable*}{hhcccchDcc}
\tablecaption{Times and Fitted Parameters of GCS Model for the CME
  \label{tab:gcs}}
\tablewidth{0pt}
\tablehead{[-1.5ex]
\nocolhead{Coronagraph} & \nocolhead{Coronagraph} & \colhead{Time} &
\colhead{Longitude} & \colhead{Longitude} & \colhead{Latitude} &
\nocolhead{Tilt angle} & \multicolumn2c{Height} & \colhead{Aspect ratio} & \colhead{Half angle}\\
\nocolhead{(\soho)} & \nocolhead{(\stereo)} & \colhead{(\stereo, UT)} &
\colhead{(Carrington, \degr)} & \colhead{(HEE, \degr)} & \colhead{(HEE, \degr)} &
\nocolhead{(\degr)} & \multicolumn2c{(\rsun)} & \colhead{} & \colhead{(\degr)}
}
\decimals
\startdata
\multicolumn{10}{l}{\soho/LASCO C2 and \stereo/COR1} \\
C2  & COR1  & 15:25  & 76.4   & 60.4 & 3   & 90  & 3.00  & 0.3  & 20\\
C2  & COR1  & 15:35  & 74.1   & 58.1 & 3   & 90  & 3.29  & 0.3  & 20\\
C2  & COR1  & 15:45  & 72.2   & 56.4 & 3   & 90  & 3.59  & 0.3  & 20\\
C2  & COR1  & 16:00  & 71.0   & 55.3 & 3   & 90  & 4.00  & 0.3  & 25\\
C2  & COR1  & 16:10  & 69.9   & 54.3 & 3   & 90  & 4.41  & 0.3  & 25\\
C2  & COR1  & 16:25  & 68.0   & 52.5 & 3   & 90  & 4.86  & 0.3  & 27\\
\multicolumn{10}{l}{\soho/LASCO C2 and \stereo/COR2} \\
C2  & COR2  & 16:39  & 64.9   & 49.5 & 2   & 90  & 5.57  & 0.4  & 30\\
C2  & COR2  & 16:54  & 63.3   & 48.1 & 2   & 90  & 6.15  & 0.4  & 30\\
C2  & COR2  & 17:24  & 62.9   & 48.0 & 2   & 90  & 7.35  & 0.4  & 30\\
\multicolumn{10}{l}{\soho/LASCO C3 and \stereo/COR2} \\
C3  & COR2  & 17:39  & 62.9   & 48.1 & 1   & 90  & 7.79  & 0.4  & 30\\
C3  & COR2  & 17:54  & 62.7   & 48.1 & 0   & 90  & 8.40  & 0.4  & 30\\
C3  & COR2  & 18:24  & 62.5   & 48.1 & 0   & 90  & 9.71  & 0.4  & 33\\
C3  & COR2  & 18:39  & 62.3   & 48.1 & 0   & 90  & 10.28 & 0.4  & 33\\
C3  & COR2  & 18:54  & 62.2   & 48.1 & 0   & 90  & 10.79 & 0.4  & 33\\
C3  & COR2  & 19:24  & 60.9   & 47.1 & 0   & 90  & 11.81 & 0.4  & 35\\
C3  & COR2  & 19:39  & 60.7   & 47.0 & 0   & 90  & 12.71 & 0.4  & 35\\
C3  & COR2  & 19:54  & 60.7   & 47.1 & 0   & 90  & 13.56 & 0.4  & 35\\
C3  & COR2  & 20:24  & 59.8   & 46.5 & 0   & 90  & 14.85 & 0.4  & 35\\
C3  & COR2  & 20:39  & 59.3   & 46.1 & 0   & 90  & 15.64 & 0.4  & 35\\
C3  & COR2  & 21:24  & 58.6   & 45.8 & 0   & 90  & 17.57 & 0.4  & 35\\
\enddata
\tablecomments{The times are from the \sta{} data.
For each time only the longitude and height are fitted,
and the other parameters are fixed and merely adjusted according to the times.
The tilt angle is set to 90\degr{} (see the text), which is not shown in the table.}
\end{deluxetable*}

\clearpage
\begin{figure}
\epsscale{.70}
\plotone{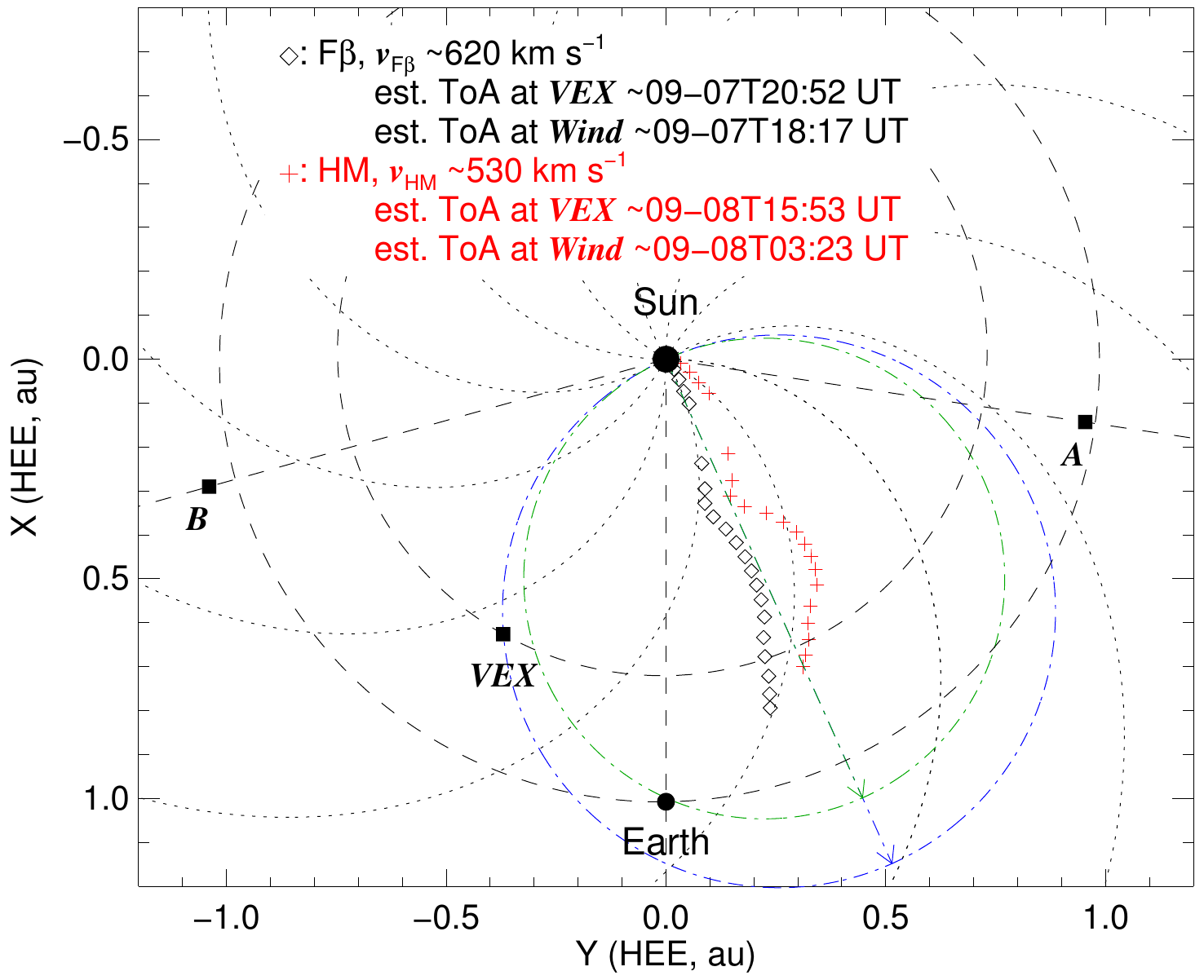}
\caption{\label{fig:pos}Positions of {\vex{}, \sta{}, \stb{}} and the Earth
in the ecliptic plane on 2010 September 8.
The trajectories of {the apex of the CME}
are derived from the triangulation method
with the \fixb{} (black diamond)
and HM (red cross) approximations, respectively.
The green (blue) circle represents the size
of the assumed spherical CME front when arriving at {\wind} (\vex),
and the green (blue) arrow indicates the direction of the CME apex then.
The black circles mark the orbits of \vex{} and the Earth, respectively,
and the gray dotted curves show Parker spiral magnetic field lines
created with a solar wind speed of 450 \kmps.
The average speeds of the CME as well as the estimated arrival times at \vex{} and \wind{}
from the two approximations are also given.}
\end{figure}

\clearpage
\begin{figure}
\epsscale{.70}
\plotone{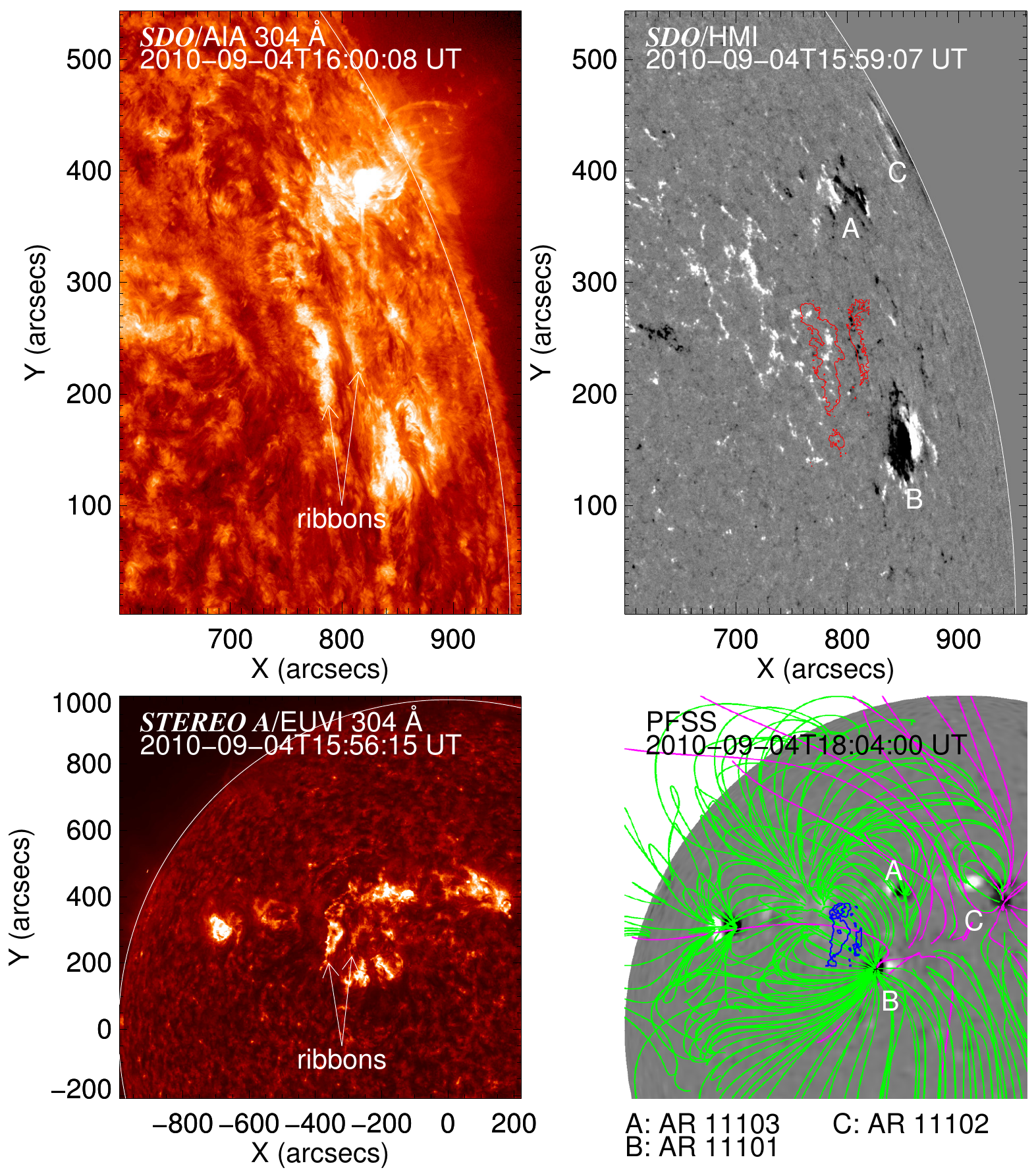}
\caption{\label{fig:euv}EUV observations and magnetic field conditions
of the CME source regions.
Upper left: EUV image at wavelength 304 {\AA} from \sdo/AIA
showing the two ribbons associated with the CME.
Upper right: Line-of-sight magnetogram from \sdo/HMI
with an overlaid AIA 304 {\AA} intensity contour of the two ribbons (red closed curves).
Lower left: EUV image at wavelength 304 {\AA} from \sta/EUVI.
Lower right: PFSS modeled magnetic field surrounding the source region
from the viewpoint of \sta,
where the green lines represent the closed field lines,
and the purple lines are the open magnetic field lines;
{the overlaid blue closed curves
are a \sta/EUVI 304 {\AA} intensity contour
indicating the positions of the two ribbons.}
The locations of the flare ribbons {are marked in the EUV images.
ARs 11103, 11101 and 11102 are indicated with ``A'', ``B'' and ``C'',
respectively, in the magnetograms.
The white (dark) regions in the magnetograms
represent the positive (negative) magnetic field regions.}}
\end{figure}

\clearpage
\begin{figure}
  \centerline{\includegraphics[scale=0.6]{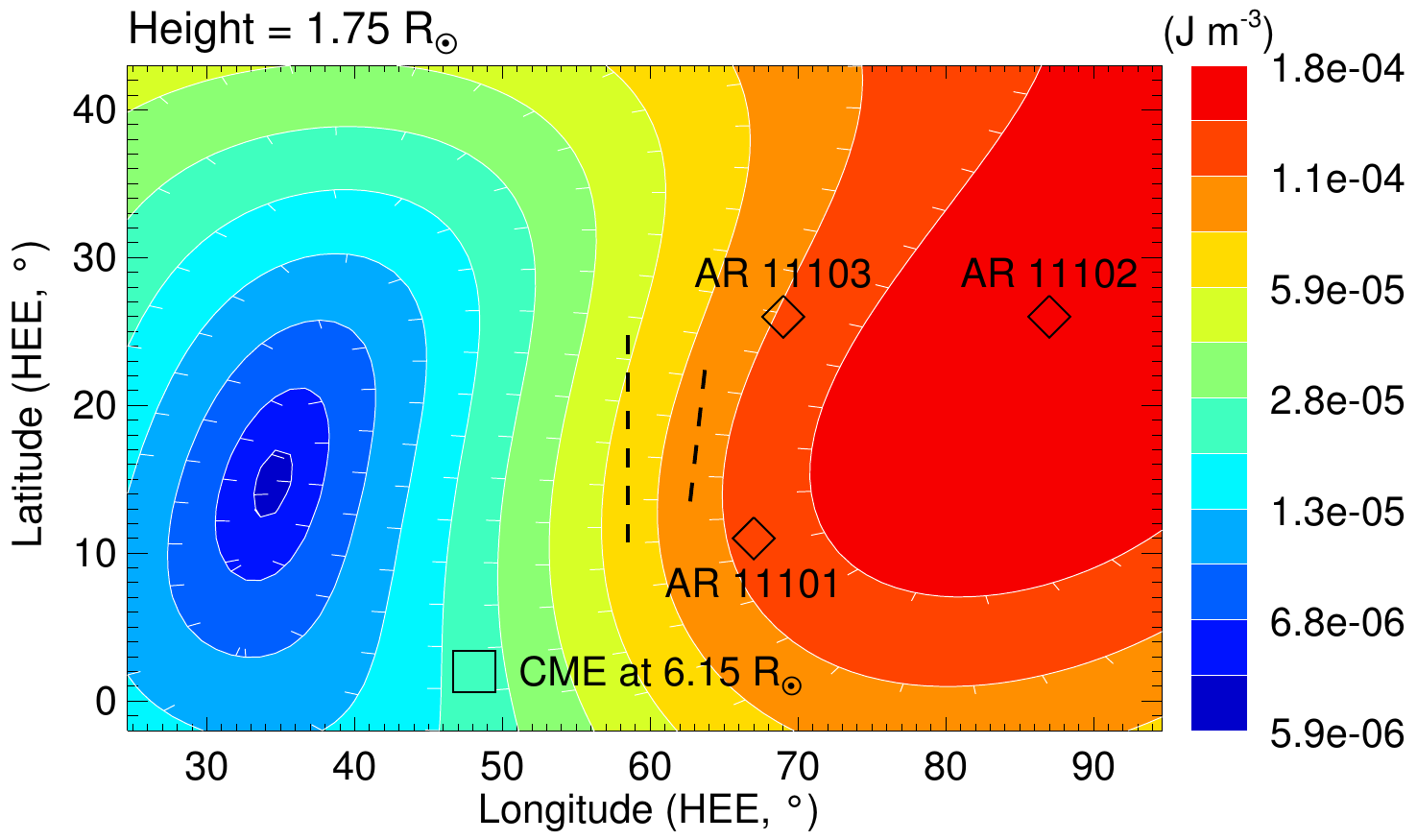}}
  \centerline{\includegraphics[scale=0.6]{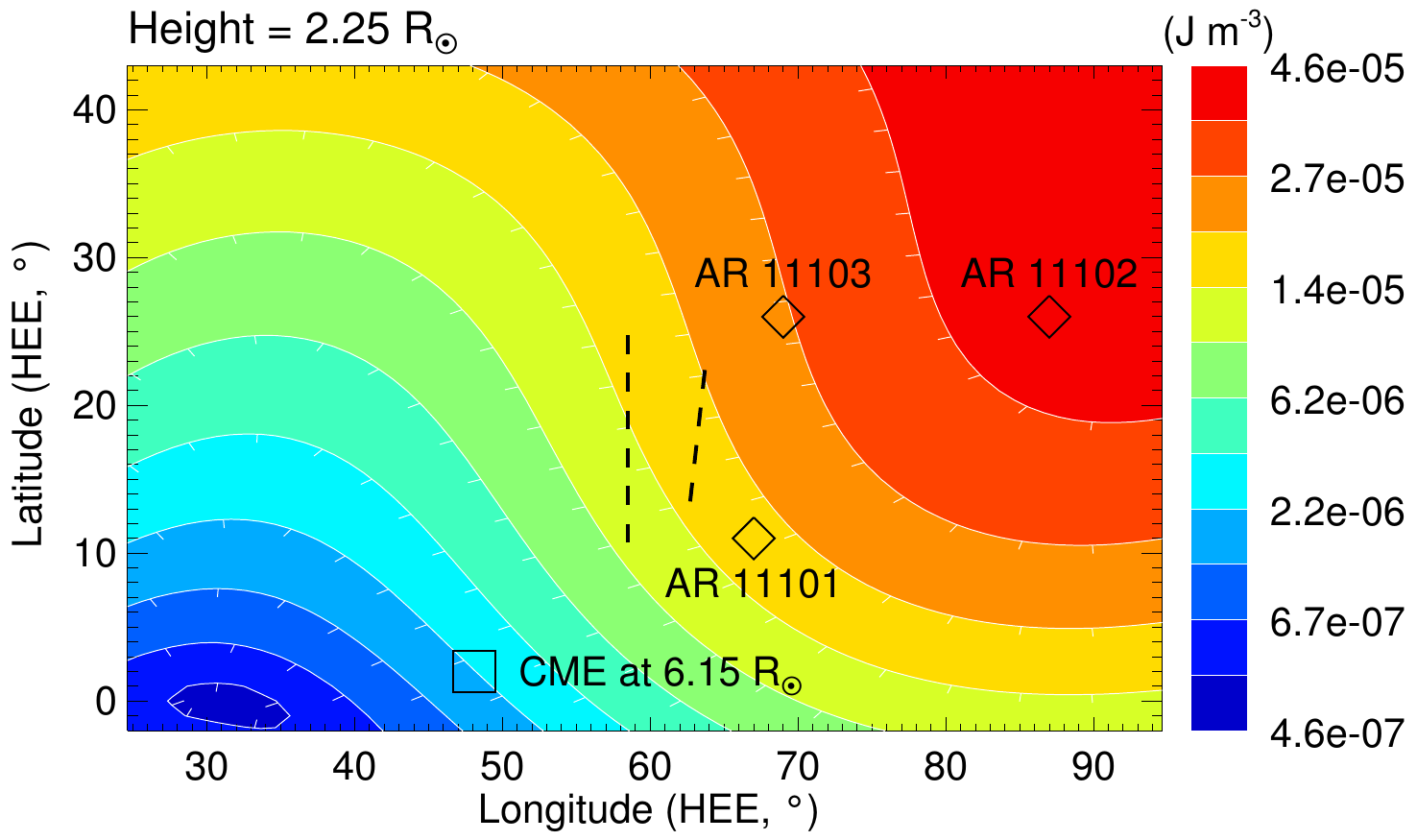}}
  \caption{\label{fig:pfss_be}{Contours of the magnetic energy density
  at heights of 1.75 \rsun{} (upper) and 2.25 \rsun{} (lower)
  from the PFSS extrapolation.
  The two vertical dashed lines indicate the approximate positions
  of the two ribbons determined from the \sta/EUVI 304 \AA{} image.
  The diamonds mark the positions of ARs 11103, 11101 and 11102, respectively.
  The square symbol represents the propagation direction at the height of 6.15 \rsun{}
  determined by the GCS model (see the text).}}
\end{figure}

\clearpage
\begin{figure}
  \centerline{\includegraphics[scale=0.6]{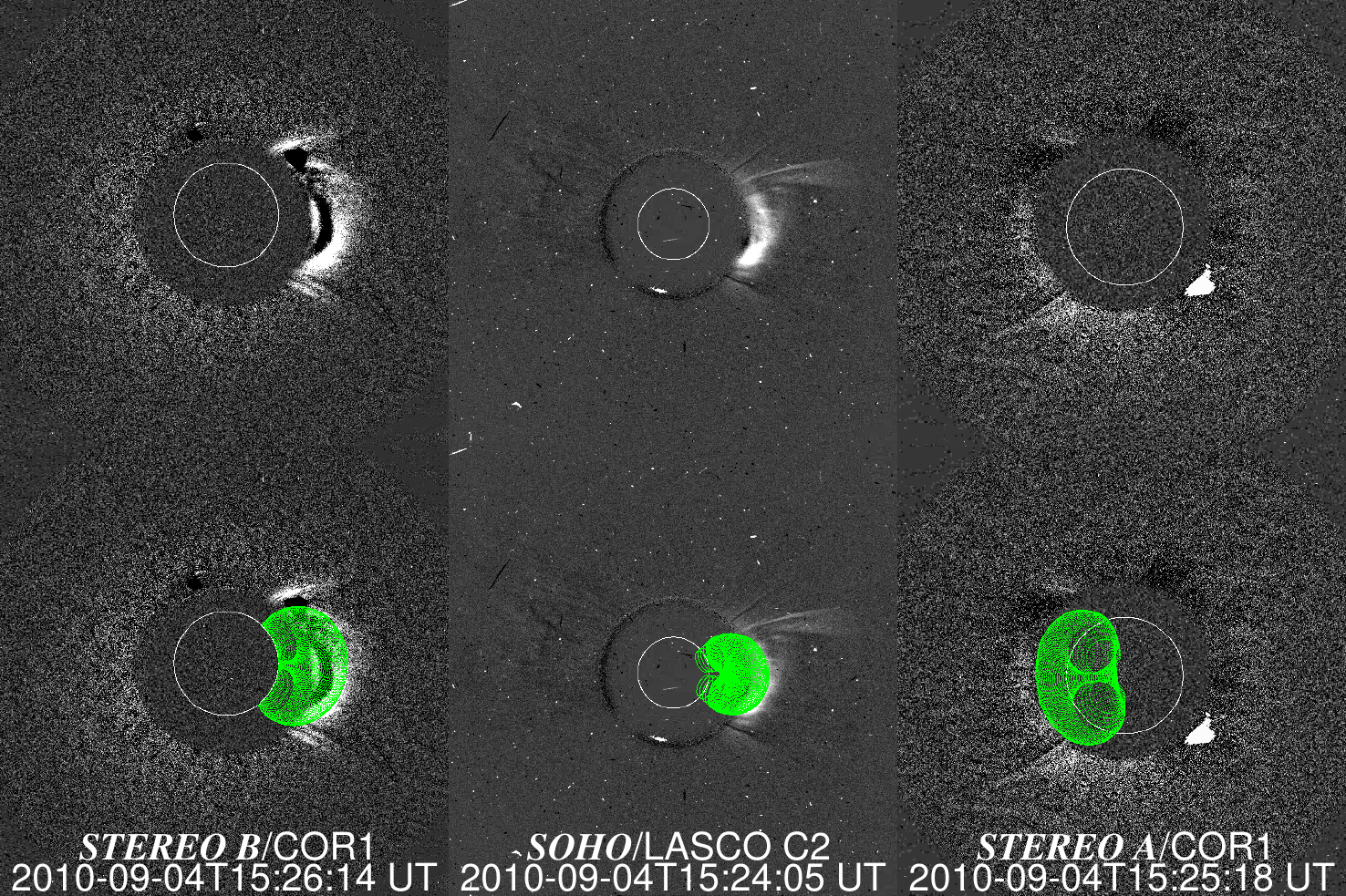}}
  \centerline{\includegraphics[scale=0.6]{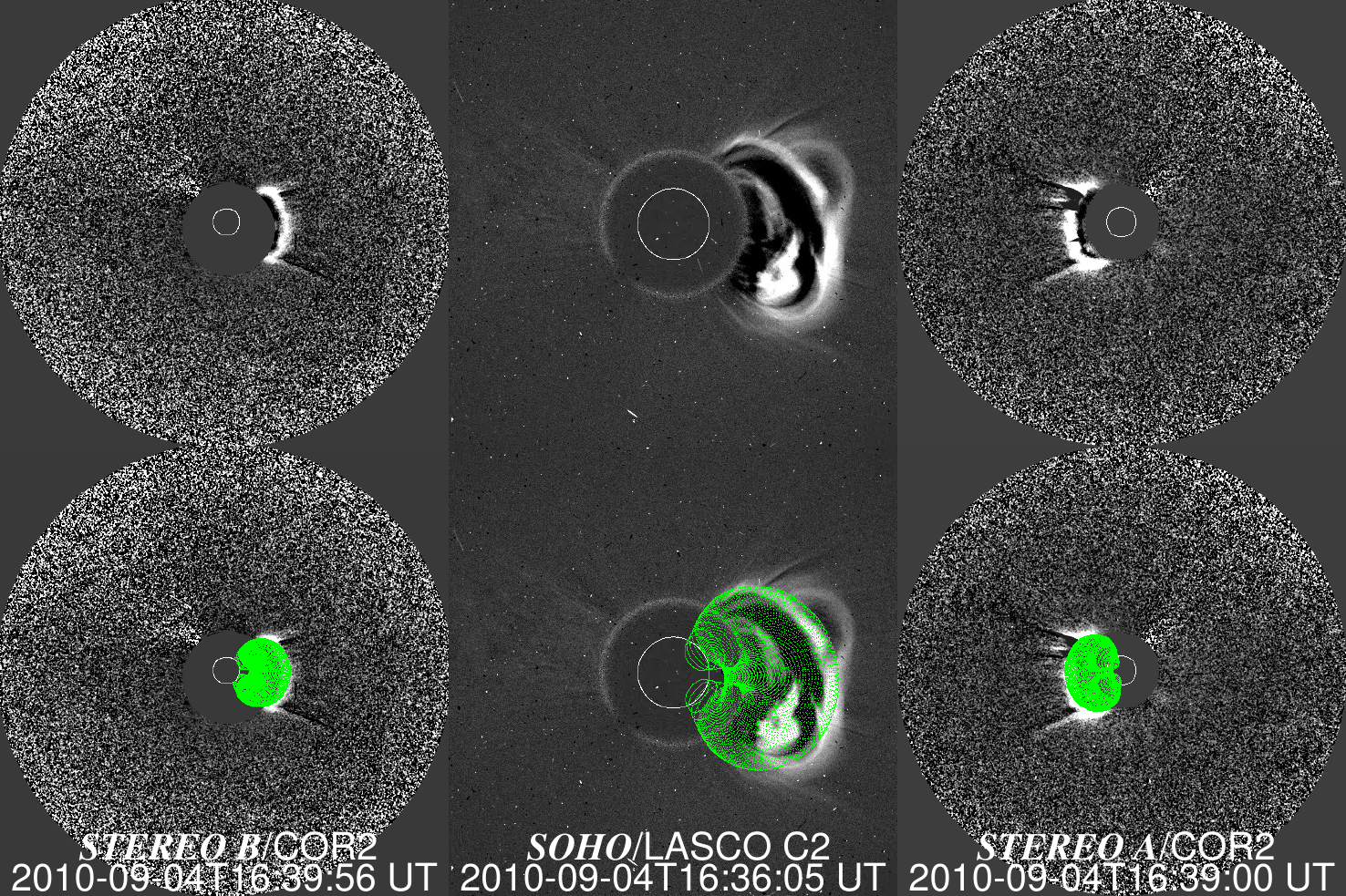}}
  \centerline{\includegraphics[scale=0.6]{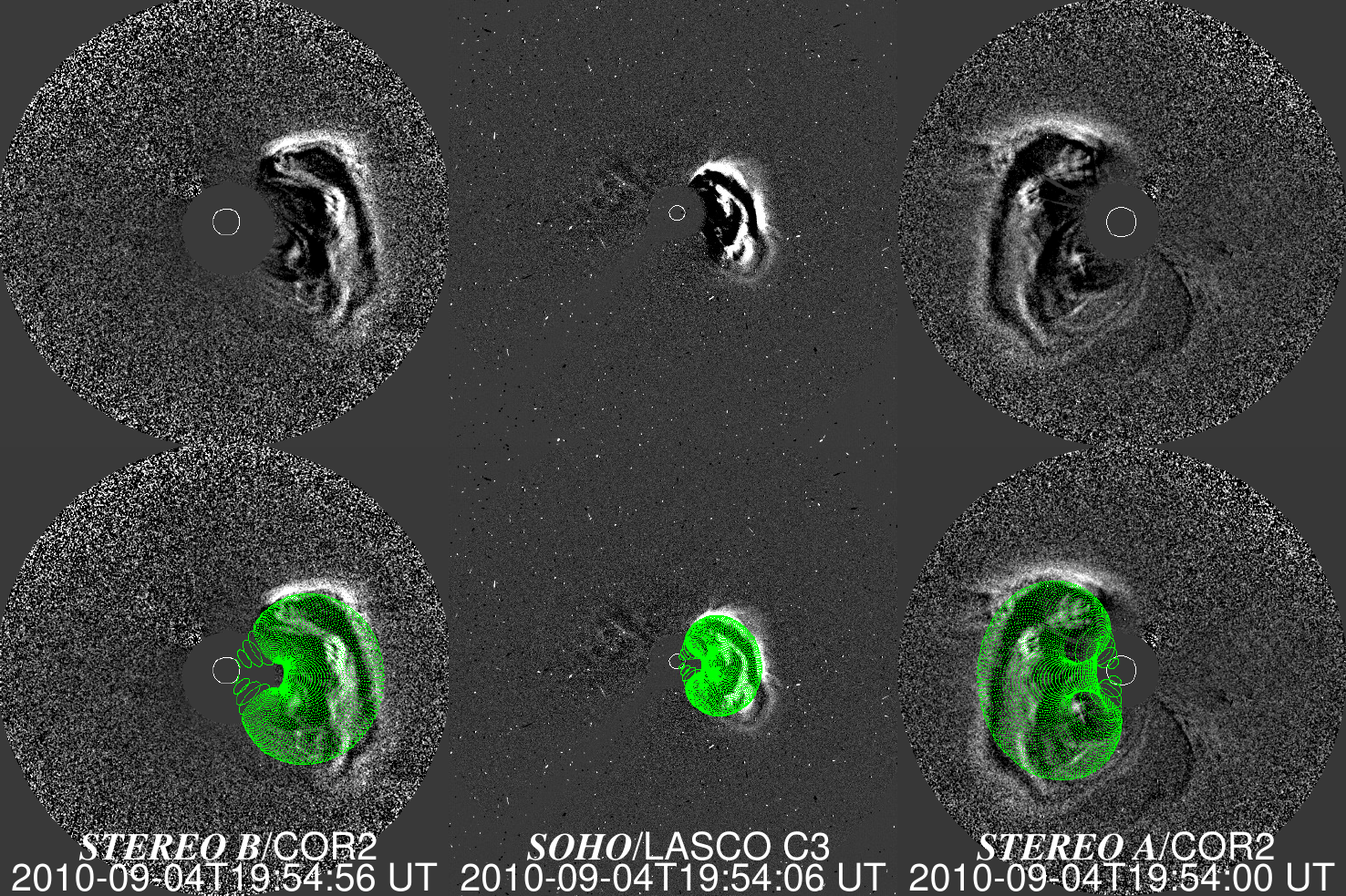}}
  \caption{\label{fig:gcs}Running difference coronagraph images
  and corresponding GCS modeling (green grids)
  from \stb{ (left column)}, \soho{ (middle column)}, and \sta{ (right column)}.
  Detectors and times are stamped in the images.}
\end{figure}

\clearpage
\begin{figure}
\plottwo{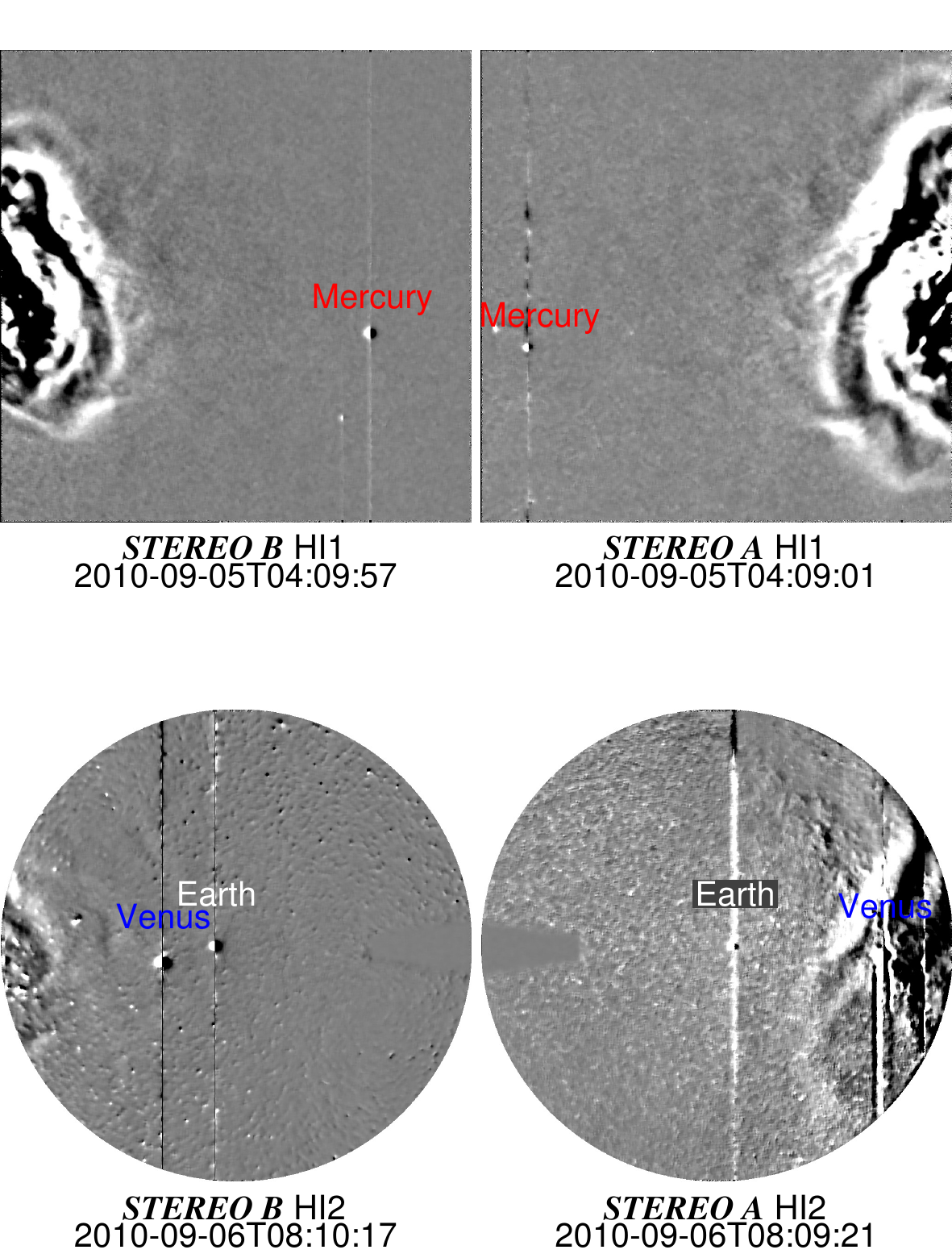}{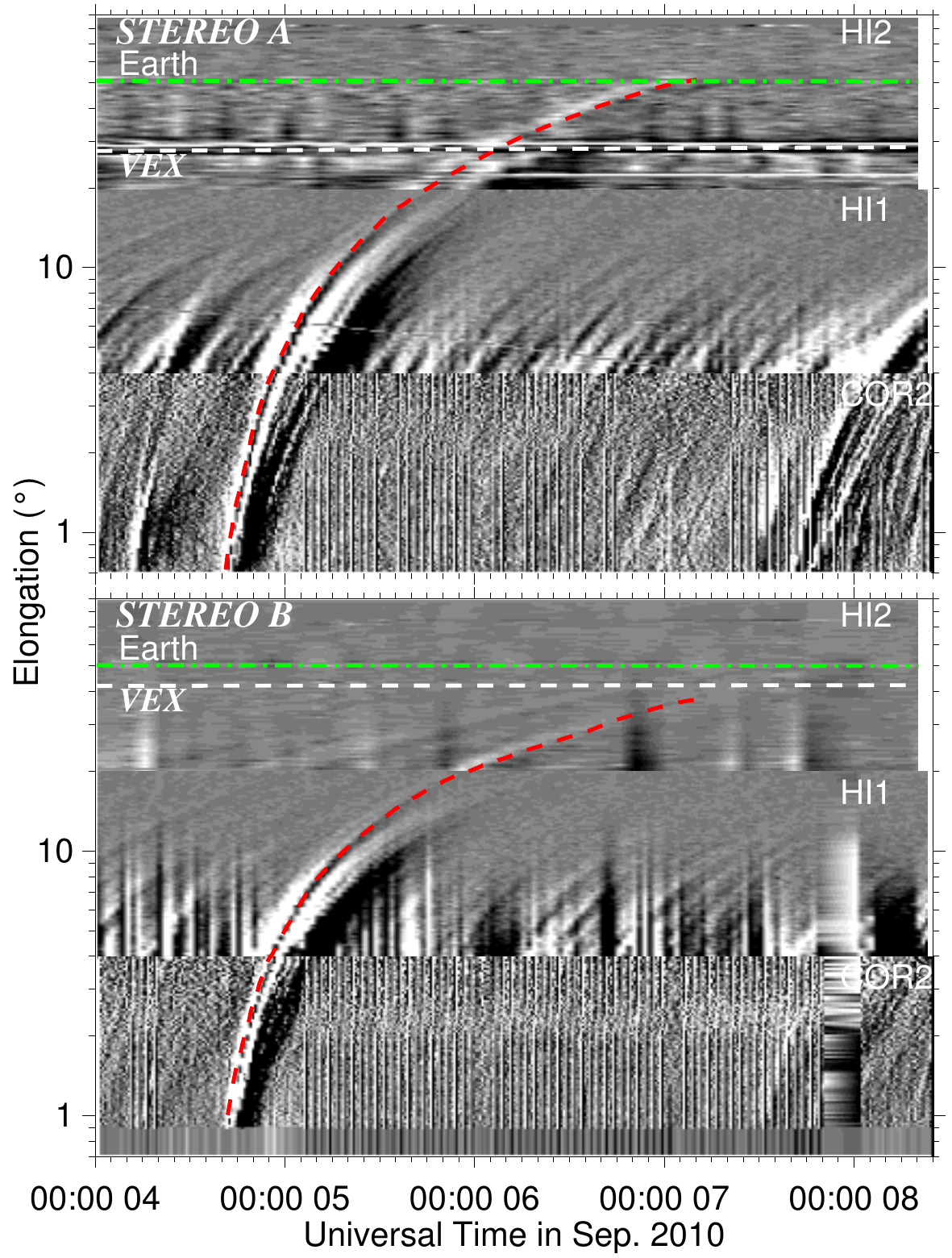}
\caption{\label{fig:hi_jmap}Left: evolution of the CME
viewed simultaneously from \stb{} and \textit{A};
the top and bottom panels show running difference images
from HI1 and HI2, respectively;
the positions of Mercury, Venus and the Earth are
marked in corresponding HI images.
Right: time-elongation maps constructed
from running difference images of COR2, HI1 and HI2
along the ecliptic plane
for \sta{} (upper) and \textit{B} (lower);
the red dashed curve indicates the track of the CME,
from which the elongation angles {of the CME front} are extracted;
the horizontal lines denote the elongation angles
of the Earth (green) and \vex{} (white).}
\end{figure}

\clearpage
\begin{figure}
\epsscale{.70}
\plotone{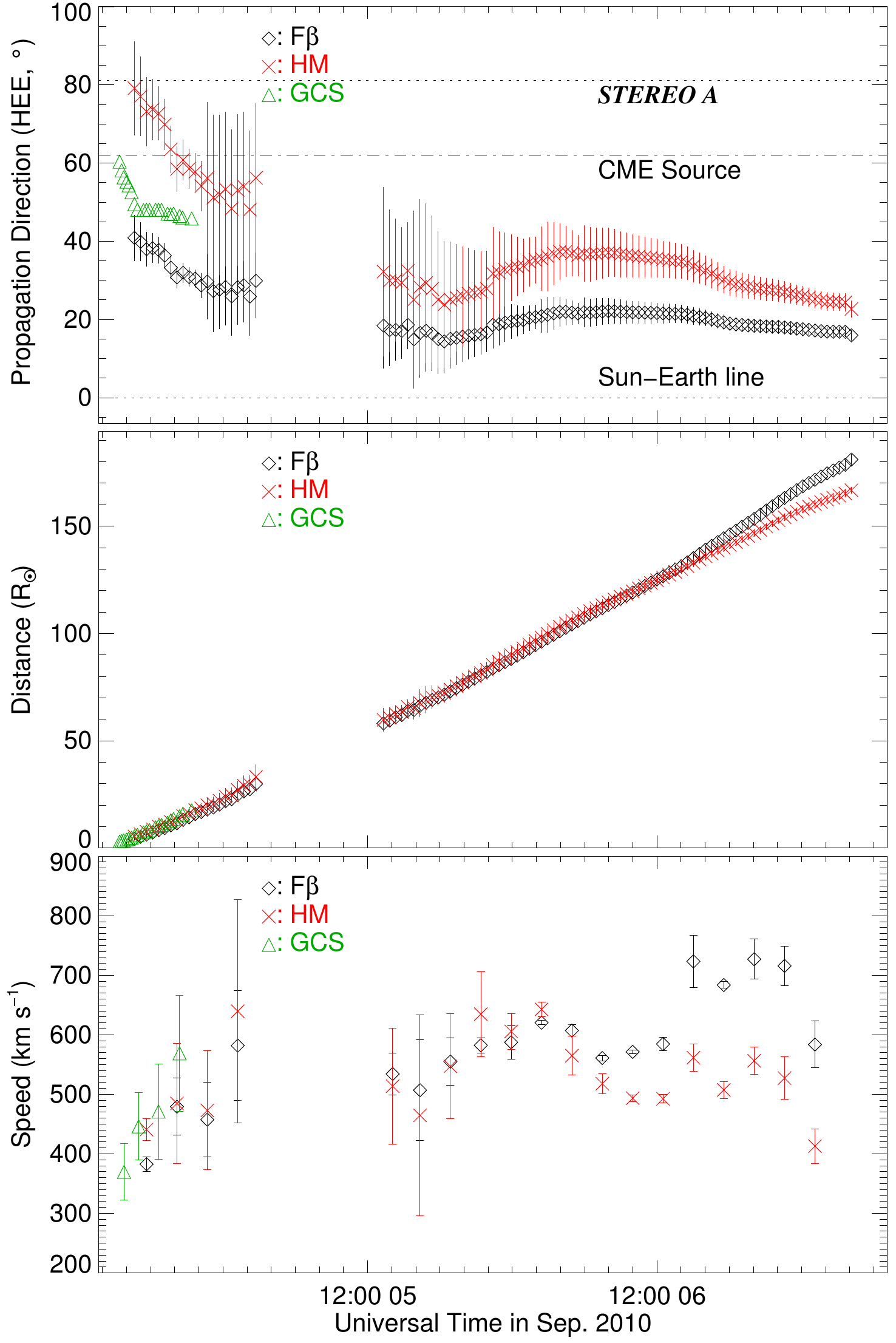}
\caption{\label{fig:prop}Propagation direction, radial distance
and speed profiles of the apex of {the CME front} derived
from the GCS model (green triangle), and the triangulation method
with the {\fixb} (black diamond) and HM (red cross) approximations.
The Sun-Earth line, and longitudes of \sta{} and the source region of the CME,
are indicated by the horizontal lines in the top panel.
The speeds are calculated from adjacent distances
using a numerical differentiation with three-point Lagrangian interpolation
and are then binned to reduce the scatter.
The data gap is due to singularities in the calculation scheme
caused by the spacecraft longitudinal separation angle
\citep{llb2011}.}
\end{figure}

\clearpage
\begin{figure}
\epsscale{.70}
\plotone{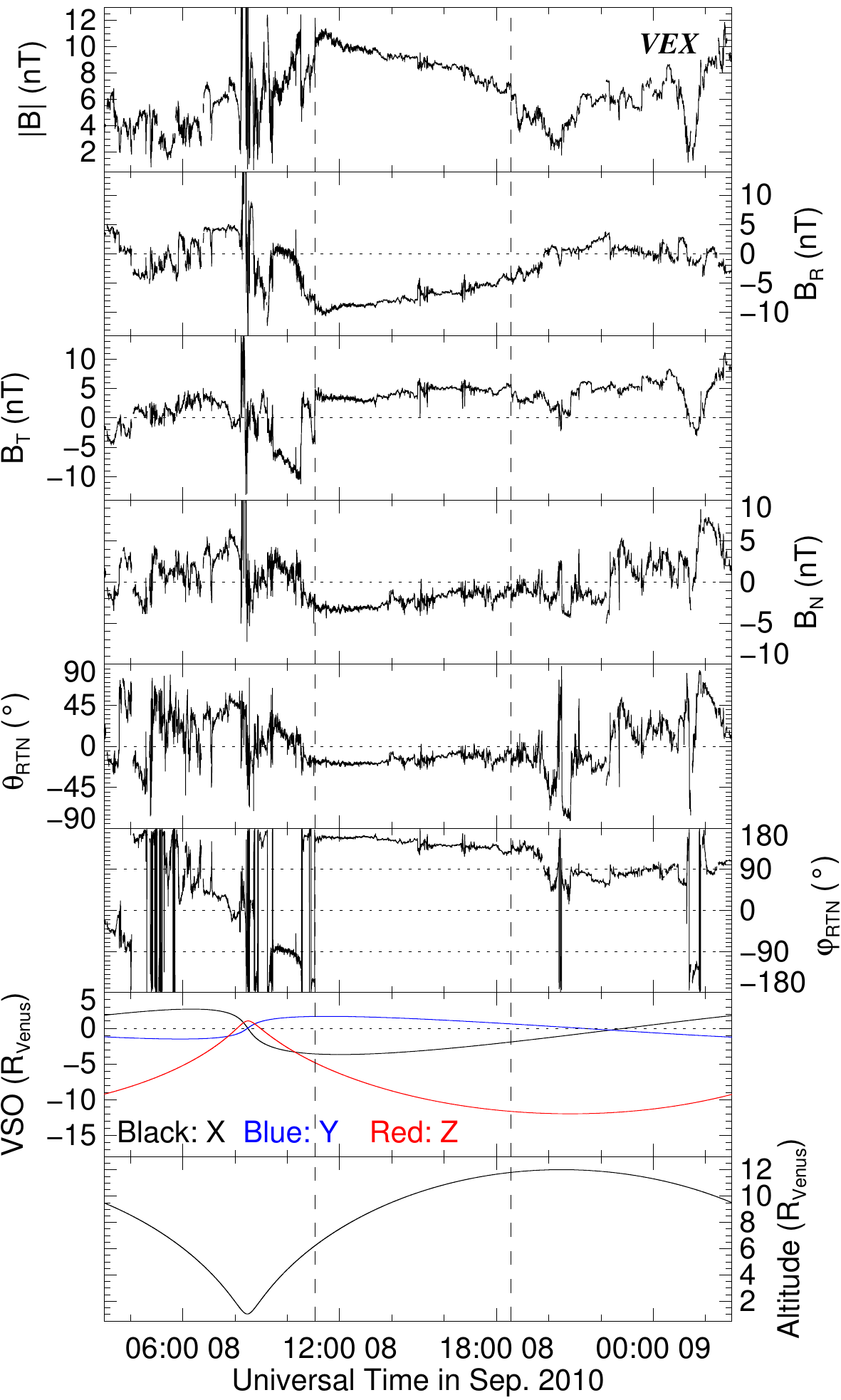}
\caption{\label{fig:vex}Magnetic field measurements at \vex{}
and the field angles in the RTN system.
Also shown are the position of the spacecraft in VSO coordinates
and the altitude from the center of Venus.
The two vertical dashed lines mark the ICME interval from
\citet{gf2016solphys}.
{The high fluctuations of the magnetic field at lower altitudes
was probably caused by the induced magnetosphere of Venus.}}
\end{figure}

\clearpage
\begin{figure}
\epsscale{.70}
\plotone{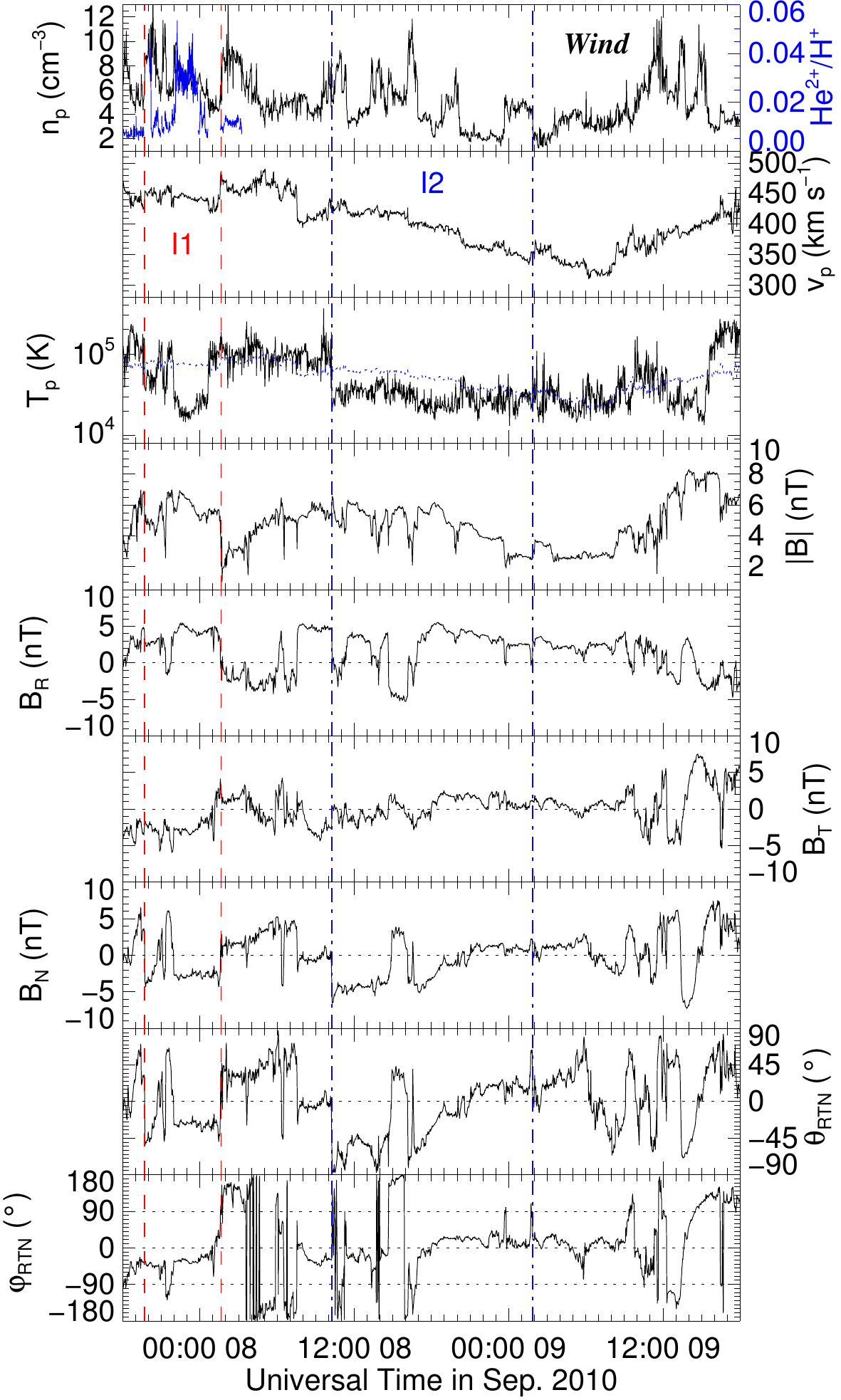}
\caption{\label{fig:wind}Solar wind plasma and magnetic field parameters
{during the event} measured at \wind.
From top to bottom, the panels show the proton density
(with the alpha-to-proton density ratio from \ace{} overlaid in blue),
bulk speed, proton temperature
(overlaid with the expected proton temperature calculated
from the observed speed
\citep{lopez1987JGR}),
magnetic field strength, components,
and field angles in the RTN system, respectively.
{The red and blue vertical lines indicate
the intervals of two ICME-like structures that are marked with
``I1'' and ``I2'', respectively.}}
\end{figure}

\clearpage
\begin{figure}
\epsscale{.70}
\plotone{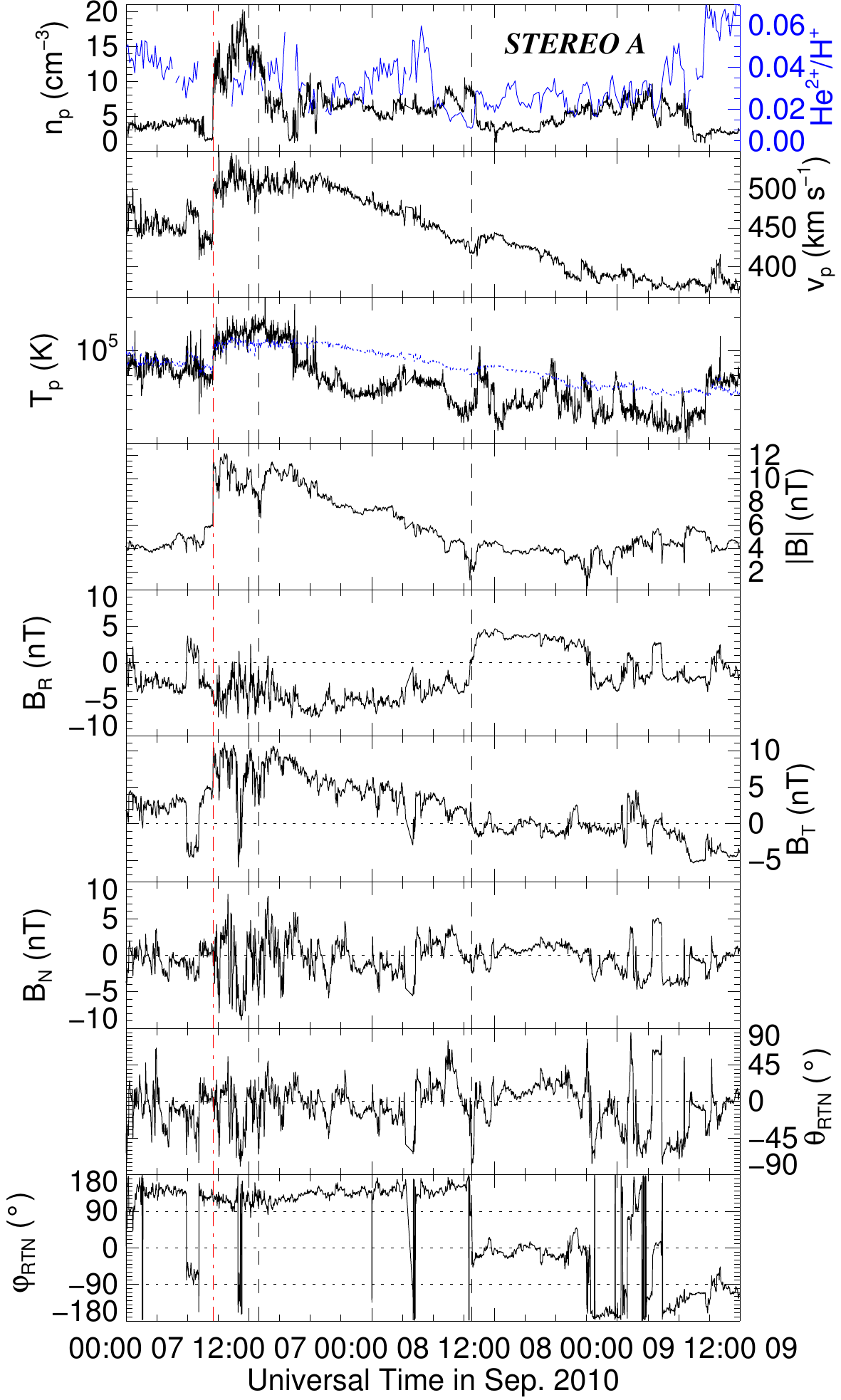}
\caption{\label{fig:sta}{An ICME driving a shock observed by \sta{}.
The panels are similar to those of Figure \ref{fig:wind}.
The red vertical line marks the arrival time of the shock,
and the two black vertical lines indicate the interval of the ICME.}}
\end{figure}

\clearpage
\bibliographystyle{aasjournal}
\bibliography{hu.cme.10sep4}

\end{document}